\documentclass[twoside,twocolumn]{article}

    \makeatletter
\def\@fnsymbol#1{\ensuremath{\ifcase#1\or 1 \or \dagger\or \ddagger\or
   \mathsection\or \mathparagraph\or \|\or **\or \dagger\dagger
   \or \ddagger\ddagger \else\@ctrerr\fi}}
    \makeatother

\usepackage[sc]{mathpazo} 
\usepackage[T1]{fontenc} 
\usepackage{microtype} 

\usepackage[english]{babel} 

\usepackage[hmarginratio=1:1,top=32mm,columnsep=20pt]{geometry} 
\usepackage[hang, small,labelfont=bf,up,textfont=it,up]{caption} 
\usepackage{booktabs} 

\usepackage{lettrine} 

\usepackage{enumitem} 
\setlist[itemize]{noitemsep} 

\usepackage{abstract} 

\usepackage{titlesec} 
\renewcommand\thesection{\Roman{section}} 
\renewcommand\thesubsection{\roman{subsection}} 
\titleformat{\section}[block]{\large\scshape\centering}{\thesection.}{1em}{} 
\titleformat{\subsection}[block]{\large}{\thesubsection.}{1em}{} 

\usepackage{fancyhdr} 
\usepackage{titling} 

\usepackage{hyperref} 

\usepackage[version=3]{mhchem} 
\usepackage[T1]{fontenc}       
\usepackage[super]{natbib}
\usepackage{physics}
\usepackage{subfig}
\usepackage{mathtools}
\usepackage{amssymb}
\usepackage{gensymb}
\usepackage{graphicx}
\usepackage{filecontents}
\usepackage{bm}
\usepackage{xr}

\usepackage{booktabs}
\usepackage{amsfonts}
\usepackage{siunitx}
\usepackage{multicol}
\pretitle{\begin{center}\Huge\bfseries} 
\posttitle{\end{center}} 
\title{Tetracycline as an inhibitor to the coronavirus SARS-CoV-2\\[2ex]}
\author{%
Tom Y. Zhao\textsuperscript{a}\thanks{To whom correspondence may be addressed. Email: tomzhao@u.northwestern.edu ; n-patankar@northwestern.edu},  
\hspace{0.05cm} Neelesh A. Patankar\textsuperscript{a1} \\[1ex] 
\normalsize \textsuperscript{a} Department of Mechanical Engineering, Northwestern University, Evanston, IL 60208 \\ 
}
\date{\small\today} 
\renewcommand{\maketitlehookd}{%
\begin{abstract}
\noindent \normalsize The coronavirus SARS-CoV-2 remains an extant threat against public health on a global scale. Cell infection begins when the spike protein of SARS-CoV-2 binds with the cell receptor, angiotensin-converting enzyme 2 (ACE2). Here, we address the role of Tetracycline as an inhibitor for the receptor-binding domain (RBD) of the spike protein. Targeted molecular investigation show that Tetracycline binds more favorably to the RBD (-9.40 kcal/mol) compared to Chloroquine (-6.31 kcal/mol) or Doxycycline (-8.08 kcal/mol) and inhibits attachment to ACE2 to a greater degree (binding efficiency of 2.98 $\frac{\text{kcal}}{\text{mol}\cdot \text{nm}^2}$ for Tetracycline-RBD, 5.59 $\frac{\text{kcal}}{\text{mol}\cdot \text{nm}^2}$ for Chloroquine-RBD, 5.16 $\frac{\text{kcal}}{\text{mol}\cdot \text{nm}^2}$ for Doxycycline-RBD). Stronger Tetracycline inhibition is verified with nonequilibrium PMF calculations, for which the Tetracycline-RBD complex exhibits the lowest free energy profile along the dissociation pathway from ACE2. Tetracycline appears to target viral residues that are usually involved in significant hydrogen bonding with ACE2; this inhibition of cellular infection complements the anti-inflammatory and cytokine suppressing capability of Tetracycline, and may further reduce the duration of ICU stays and mechanical ventilation induced by the coronavirus SARS-CoV-2.

\end{abstract}
}

\begin{document}

\maketitle

\section{Introduction}

The extreme urgency for therapeutics against the acute respiratory syndrome coronavirus 2 (SARS-CoV-2) drives the review of existing drugs for their ability to inhibit the function of this virus \cite{2020Omar,2020Amin}. 

Tetracycline has been proposed as a strong candidate against SARS-CoV-2 \cite{2020Sodhi} due to its lipophilic nature, anti-inflammatory response, as well as its ability to chelate zinc species on matrix metalloproteinases (MMPs). Tetracycline class antibiotics have also been shown to be effective in reducing the duration of ventilatory support and ICU stay from acute respiratory distress syndrome \cite{2020Traverso}, and Doxycycline has been suggested to be an important component in combination therapy for its anti-viral properties \cite{2020Kontoyiannis}. Tetracycline as well as a broad band of related antibiotics have been approved by the FDA \cite{2020Yates, 2019Andrei}. 

In this work, we quantify the performance of Tetracycline in inhibiting the binding of the SARS-CoV 2 spike protein to ACE2. Tetracycline is found to bind more favorably to the receptor binding domain (RBD) of the spike protein compared to Doxycycline or Chloroquine, which was included in this study as a baseline. The Tetracycline-RBD complex also displays lower binding efficiency to the human cell receptor ACE2. 

\section{Methods}
The SARS-CoV 2 RBD, ACE2, Tetracycline, and Chloroquine molecular structures were obtained from RCSB PDB (6M0J, 2UXO, 4V2O, 2XRL)  \cite{2020Lan, 2007Alguel, 2016Huta, 2020Palm}. Missing hydrogen atoms were appended, after which structural preparation and molecular docking with full ligand and protein backbone flexibility were carried out using the Rosetta suite \cite{2009Davis, 2012Lemmon, 2006Meiler}. The resulting complexes were inspected manually, after which the binding affinities of the best-scoring complexes were gauged using MM/PBSA calculations after 100 ns equilibrium molecular dynamics simulations \cite{1995Berendsen, 2001Lindahl, 2014Kumari}. The potentials of mean force (PMF) \cite{2011Chen} along the dissociation pathway of these RBD complexes from ACE2 were found in LAMMPS \cite{1995LAMMPS} using steered molecular dynamics after parameterization with CHARMM\cite{2009Brooks}. Jarzynski's equality was employed to calculate the free energy profile for each RBD complex from 10 statistically independent trajectories \cite{2013Dellago}.
 
 \section{Results and Discussion}
\begin{figure}[]
\centering
\includegraphics[width=8.5cm]{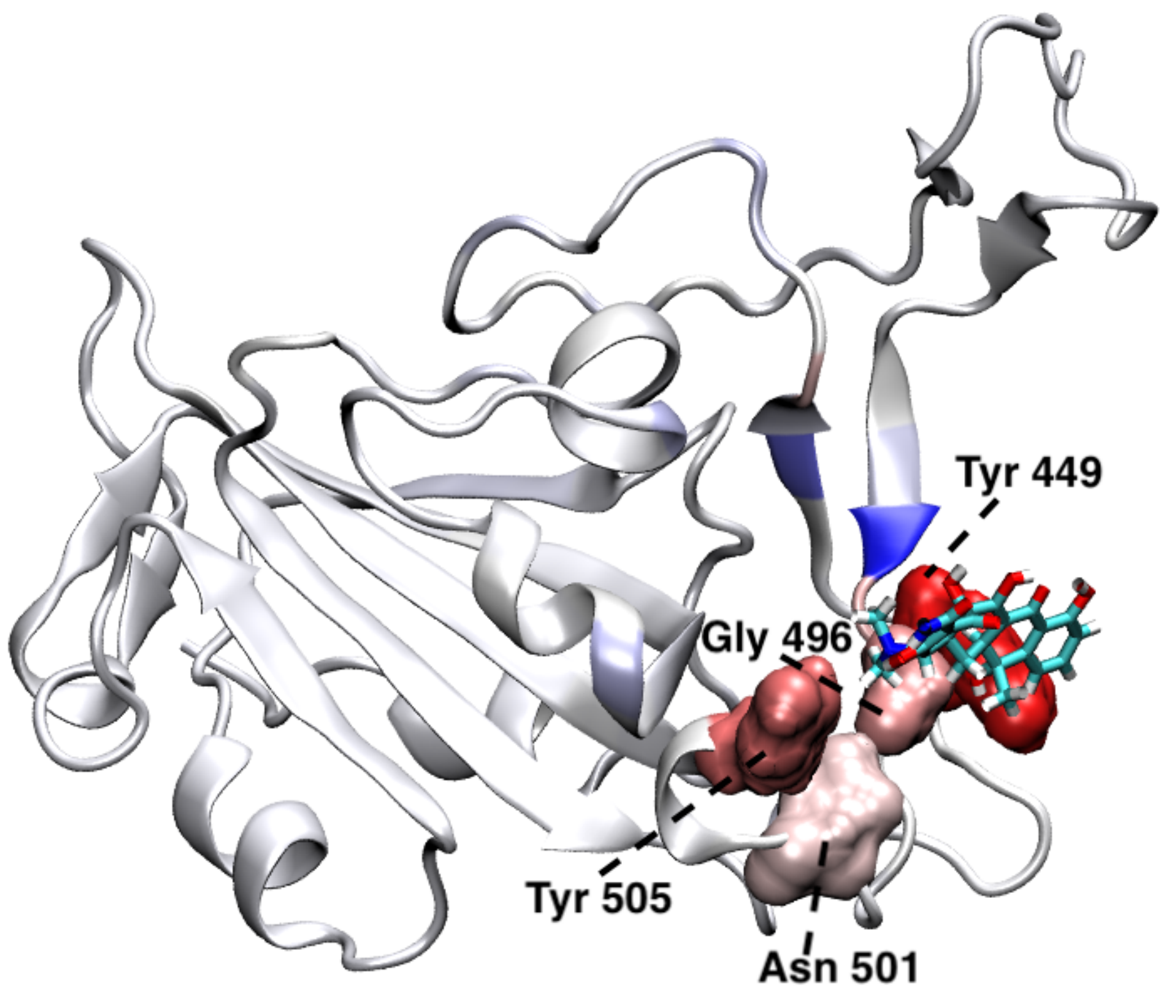}
\caption{Interaction map of amino acid residues of the SARS-CoV 2 receptor binding domain (RBD) that have the highest binding affinity with Tetracycline. The Tyr 449, Asn 501, Gly 496 and Tyr 505 residues have also been shown to form persistent hydrogen bonds in maintaining the RBD-ACE2 complex \cite{202Veer}. }
\label{fig: resmol}
\end{figure}

Tetracycline exhibits higher binding affinity to the RBD in both blind and site-specific docking (-9.40 kcal/mol) compared to Doxycycline (-8.08 kcal/mol) or Chloroquine (-6.63 vs 6.31 kcal/mol) as delineated in Table \ref{table: bindaff}. The amino acid residues of the RBD involved in hydrogen bonding with the Tetracycline molecule are Tyr 449, Asn 501, Gly 496, and Tyr 505 (Fig. \ref{fig: resmol}), which have been shown to be crucial for the SARS-CoV 2 RBD in binding to ACE2 for cellular access \cite{202Veer}. These four residues comprise major hot spots that form persistent hydrogen bonds with ACE2. Meanwhile, the amino acids of RBD that interact with Chloroquine in the site-specific configuration are Lys 356, Arg 454, Arg 466 and Arg 355, of which none are involved in extended hydrogen bonding with ACE2. 

Tetracyline appears to bind preferably to polar or slightly lipophilic RBD residues, which comprise the majority of amino acids that form persistent hydrogen bonds with ACE2\cite{202Veer, 2006Du}. Other tetracycline derivatives as Doxycycline or Minocycline are known to be more lipophilic \cite{2020Sodhi, 2020Yates, 2006Du} and may therefore prefer nonpolar residues \cite{2019Gautieri} that are often buried beneath the solvent accessible surface area of the spike protein. Indeed, the RBD residues that have highest binding affinity to Doxycycline are Tyr 449, Gly 447, Val 445, Gly 496, of which only two overlap with RBD amino acids that engage in extended hydrogen bonding with ACE2. On the other end of the spectrum, Chloroquine targets clusters of charged residues on the RBD that do not actively participate in hydrogen bonding with the cell receptor ACE2.  

\begin{table}
\noindent\begin{tabular}{@{\hskip -0.01cm}SSS@{}} \toprule
{Inhibitor} & {$\Delta G_{\text{bind}}$ (kcal/mol)} & {$\Delta G_{\text{bind}}$ (kcal/mol)}\\ 
    {to RBD}  &  {\it{This work}}    & {\it{Prior literature}}  \\ \midrule
    {Tetracycline} & -9.40 & {N/A} \\
    {Doxycycline} & -8.08 & {N/A} \\
    {Chloroquine}  & -6.31  & {[-5.3, -7.1]}  \cite{2020Nim, 2020Rohan} \\
 \end{tabular}
 \caption{The binding free energy of small-molecule inhibitors to the SARS-CoV2 receptor binding domain (RBD). Tetracycline binds preferably to the RBD.}
 \label{table: bindaff}
 \end{table}

\begin{figure}[]
\centering
\includegraphics[width=8.5cm]{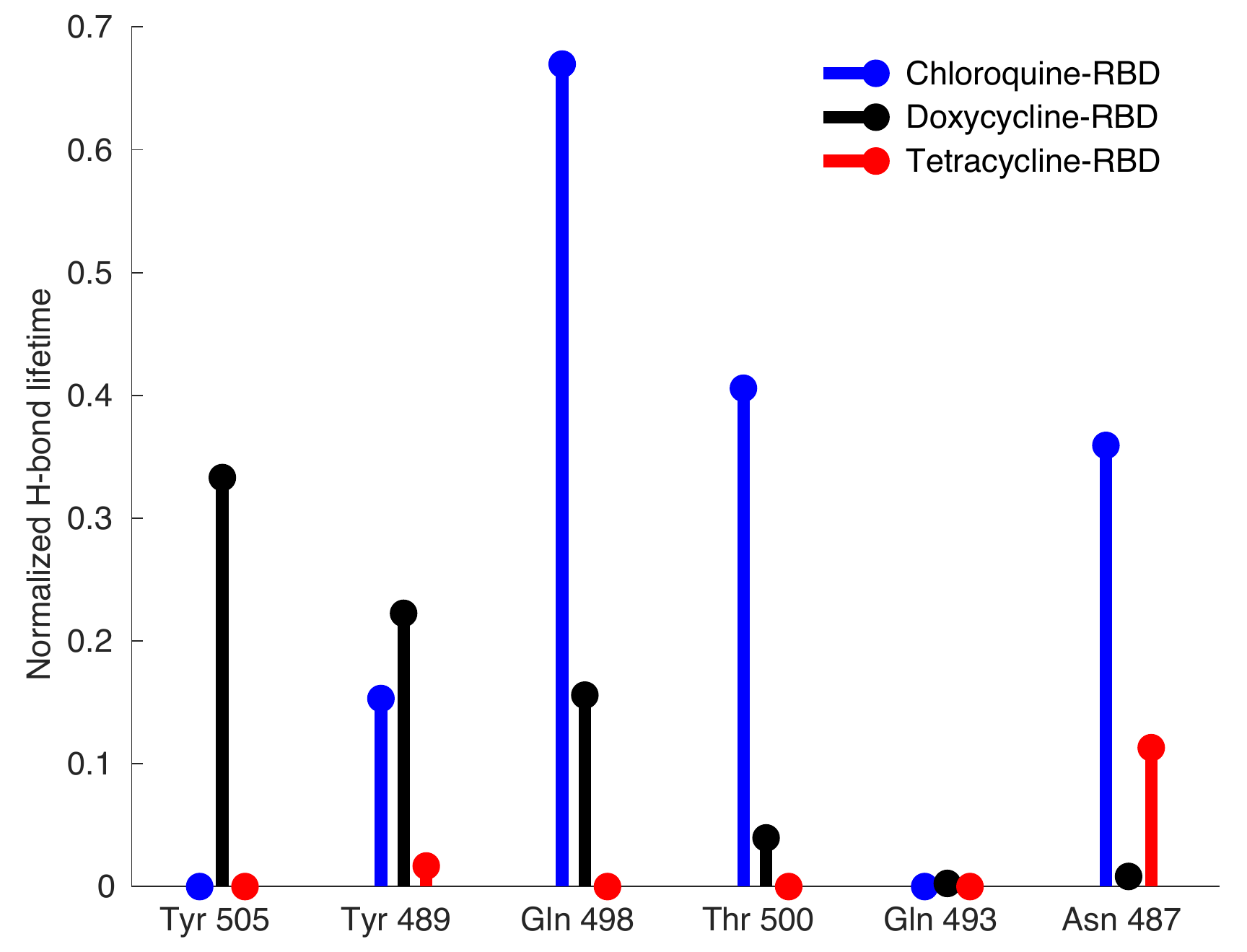}
\caption{The hydrogen bonding lifetimes of binding site residues of the inhibited RBD with ACE2 sustained in $100$ ns of simulation time, normalized by hydrogen bonding lifetimes in the uninhibited RBD-ACE2 complex.}
\label{fig: resact}
\end{figure}

The binding efficiency\cite{2012Day} (magnitude of binding energy normalized by contact interface area) of the SARS-CoV2 RBD-ACE2 complex was found to be 7.58 kcal/(mol$\cdot$nm$^2$). In the presence of the protein-ligand complex Tetracycline-RBD, the binding efficiency with ACE2 (2.98 kcal/(mol$\cdot$nm$^2$)) is significantly lower than that for Chloroquine-RBD (5.59  kcal/(mol$\cdot$nm$^2$)) and Doxycycline-RBD (5.16 kcal/(mol$\cdot$nm$^2$)) as displayed in Table \ref{table: bindaffA}. A survey of hydrogen bonding lifetimes between the important binding site residues in the RBD with ACE2\cite{202Veer} shows that the Tetracycline inhibited RBD exhibits the least hydrogen bonding activity with ACE2 (Fig. \ref{fig: resact}). This suggests that not only does Tetracycline bind more favorably to the receptor binding domain of the spike protein, it also inhibits the binding of the RBD to ACE2 to a greater degree. 

\begin{figure}[]
\centering
\includegraphics[width=8.5cm]{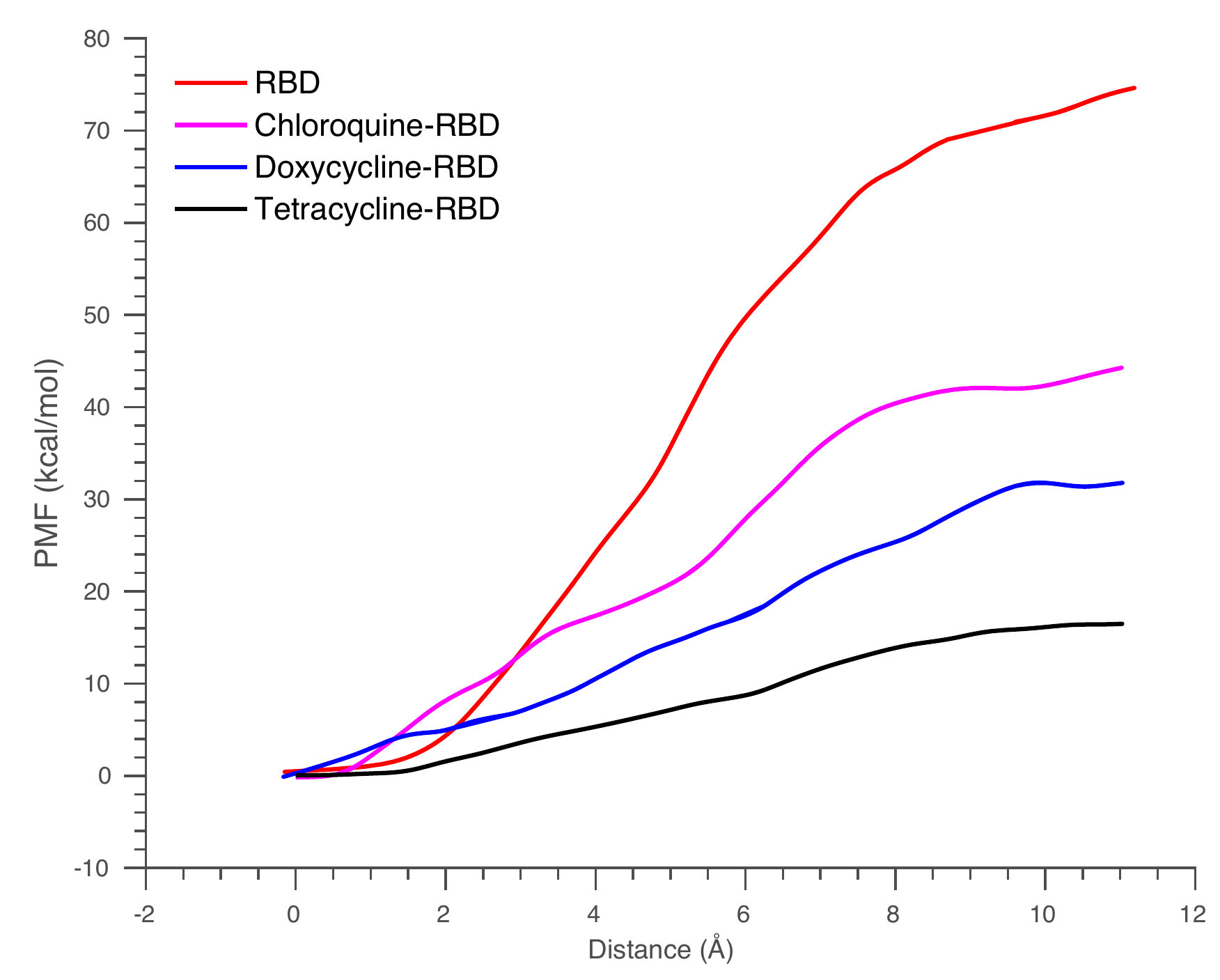}
\caption{The potential of mean force (PMF) as a function of the distance between the centers of masses of the spike protein RBD complexes and the cell receptor ACE2. The Tetracycline-RBD complex exhibits the lowest free energy profile along the dissociation pathway.}
\label{fig: pmf}
\end{figure}

To verify this statement, steered molecular dynamics simulations were carried out to find the potential of mean force (PMF) along a singular dissociation pathway for the inhibited and uninhibited RBD-ACE2 complexes. Figure \ref{fig: pmf} shows that the PMF for unbinding of the Tetracycline-RBD complex from ACE2 was lowest of the three structures tested, which is in agreement with the binding efficiencies found from equilibrium simulations. This disruption of the RBD-ACE2 interface may therefore inhibit the signaling cascade initiated during binding of the viral spike protein.
\begin{table}
\noindent\begin{tabular}{@{\hskip -0.01cm}SSS@{}} \toprule
{Complex} & {Binding efficiency} & {}\\ 
    {to ACE2}  &  {(kcal/(mol$\cdot$nm$^2$))}     \\ \midrule
    {RBD} & 7.58  \\ 
    {Chloroquine-RBD} & 5.59   \\
    {Doxycycline-RBD} & 5.16 \\
        {Tetracycline-RBD}  & 2.98  \\ \midrule
 \end{tabular}
 \caption{The binding efficiency\cite{2012Day} (magnitude of binding energy normalized by contact interface area) of the spike protein RBD as well as the Tetracycline-RBD, Doxycycline-RBD and Chloroquine-RBD complexes to the human cell receptor ACE2. Binding efficiency is lowest for the Tetracycline-RBD complex, indicating that Tetracycline is a more effective inhibitor.}
 \label{table: bindaffA}
 \end{table}
\section{Conclusion}
The tetracycline class of antibiotics, including Tetracycline, Oxytetracycline, and Doxycycline may be helpful in the fight against the coronavirus SARS-CoV-2, due to its preferential association with the important residues in the viral receptor binding domain and the resulting strong inhibition of the RBD-ACE2 complex. Further experimental studies are recommended to validate how this reduction of cellular infection complements or enhances the anti-inflammatory and anti-viral properties of tetracyclines in their role as treatment for SARS-CoV-2.


\section*{Author contributions}
T.Y.Z conceived and planned the research, as well as performed calculations. N.A.P. and T.Y.Z. performed analysis and wrote the manuscript.
\section*{Competing interests}
The authors have no competing financial interests or other interests that might be perceived to influence the results and/or discussion reported in this paper.

\newpage

\providecommand{\latin}[1]{#1}
\makeatletter
\providecommand{\doi}
  {\begingroup\let\do\@makeother\dospecials
  \catcode`\{=1 \catcode`\}=2 \doi@aux}
\providecommand{\doi@aux}[1]{\endgroup\texttt{#1}}
\makeatother
\providecommand*\mcitethebibliography{\thebibliography}
\csname @ifundefined\endcsname{endmcitethebibliography}
  {\let\endmcitethebibliography\endthebibliography}{}

%
%
%


\end{document}